\documentstyle[11pt]{article}
\input{psfig}
\begin{document}
\vspace{5mm}
\begin{flushright}
Liverpool Preprint LTH 344 \\
Swansea Preprint SWAT/58 \\
hep-lat/9412079 \\
\end{flushright}
\begin{center}
{\LARGE\bf
Efficient Hadronic Operators in Lattice Gauge Theory
}\\[10mm] {\large\it UKQCD Collaboration}\\[3mm]

{\bf  P. Lacock, A. McKerrell, C. Michael, I. M. Stopher}\\

{DAMTP, University of Liverpool, Liverpool, L69 3BX, U.K.}

{\bf and}\\

{\bf  P. W. Stephenson}\\

{Department of Physics, University College of Swansea, Swansea, %
SA2 8PP, U.K.}

\end{center}

\begin{abstract}

We study operators to create hadronic states made of light  quarks in
quenched lattice gauge theory.  We construct non-local gauge-invariant
operators which  provide information  about the spatial extent of the
ground state and excited states.  The efficiency of the operators is
shown by looking at the wave function of the first excited state, which
has a node as a  function of the spatial extent of the operator. This
allows one  to obtain an uncontaminated ground state for hadrons. 

\end{abstract}

\section{Introduction}

In quenched lattice gauge theory, hadronic states are created by acting
with light quark (or anti-quark) creation operators on  the vacuum. The
propagation of such a state in (euclidean) time then allows  its mass
to be determined. Provided sufficiently large euclidean time $t$ is 
taken, the mass determination is independent of the particular hadronic
creation operator used, since the ground state always dominates.  In 
practice, a limited range of $t$ is available, so methods are needed to
create the  hadronic ground state efficiently. For a study of hadronic
matrix elements, the requirements for an efficient hadronic operator 
are even stronger since off-diagonal terms involving excited states
have to be removed. This has been a topic  of much study --- smearing the
quark propagators has been used for example~\cite{smear1,smear2}.

The main contamination of the ground state signal at larger $t$ values 
comes from the first excited state. So, effectively, one requires a 
hadronic creation operator which maximises the ground state relative 
to this first excited state. We study this explicitly by making 
simultaneous two-exponential fits to various hadron operators.

This relative amplitude is usually called the Bethe Saltpeter (or BS)
wave function of the hadron. It is the overlap between  a quark and 
antiquark at distance $R$ apart and the hadronic state which is an 
eigenstate of the hamiltonian (transfer matrix on a lattice). We use  a
fuzzed gluon flux prescription~\cite{bs-gup} to join the quarks in a
gauge-invariant way. This corresponds to  the adiabatic wave function
as defined by~\cite{tn}.  Some previous  work has used quark and
antiquarks in the (spatial) coulomb gauge  instead~\cite{bs-gup,tn}. 
This is less efficient (in our sense) and also leads to problems with 
image sources in the spatial periodic boundary conditions.

The paper is organised as follows: in the next section we  construct
non-local gauge-invariant operators using fuzzed links. In Sec. 3 we 
apply these ideas to the $\pi$ and $\rho$ mesons and to baryons. We
explicitly fit the correlation between hadronic operators at  time
separation $t$ using both ground state and excited states. Details about
the lattices and couplings are also given. In Sec. 4 we investigate the
BS wave function obtained  from the fit analysis. We show that the wave
function has  a node as a function of flux tube length,  which points to
the value of $R$ to be used for an efficient  operator to create the
ground state.
 
\section{Effective Operators}

The idea of applying gauge-invariant smearing techniques to the
propagator at the source and/or sink to enhance the overlap with the
ground state has been used extensively in the
literature~\cite{smear1,smear2}. These methods rely on a smearing 
derived from a scalar massive particle propagating in the spatial 
dimensions only. There is no direct physical motivation for  such an
assumption, although it does work in practice.

One clear motivation for a trial hadronic operator comes from 
consideration of the well-understood case of heavy-quark hadrons. For 
mesons, the $c\overline{c}$ and $b\overline{b}$ states are approximately
non-relativistic. The adiabatic approximation is also well justified, so
that they can be modelled as heavy point particles (quarks) bound  by a
central potential $V(R)$  between static colour sources (in the
fundamental colour representation). On a lattice, efficient gluonic
operators which create such a colour flux  between static sources are
known. These operators are constructed using a fuzzing algorithm, which
can be implemented recursively~\cite{smear}. This iterative fuzzing
prescription  creates gluonic fluxes with a dominant ground state and
only a small admixture of excited states. This is the standard way to
measure the potential energy $V(R)$ accurately in lattice QCD.   To
extend this approach  to lighter quarks, we then use such a lattice
construction of a colour   flux tube of length $R$ to join two light
quarks in a gauge invariant  manner. Then we may vary $R$ and explore
the relative amplitude of  ground state and excited state hadron
created.

This method has already been explored~\cite{bs-gup}.  Here we construct
fuzzed gluon flux tubes following exactly the successful methods  used
for studying the potential~\cite{smear}. An iterative  fuzzing is used,
summing over spatial u-bends with three links:
 \begin{equation}
 U_{new} = P_{SU(3)} (c U_{old} + \sum_1^4 U_{u-bends} )
\label{eqf}
 \end{equation}
 where a projection of a matrix $M$ to the SU(3) matrix $U$ is carried
out iteratively by maximising  ${\rm Re}\;{\rm Trace}\;(M U^{\dag})$
using a Cabibbo-Marinari approach.  The best overlap with the ground
state  potential $V(R)$ comes from large $c$ and large number of
iterative  steps~\cite{ukqcdpot}. However, in the present study, varying
the parameters of the fuzzing  prescription gave relatively little
change. Hence we select a smaller fuzzing level (5)  and coarser fuzzing
($c=2$) to minimise computer resources.

The motivation of the method we are using is to join the quark  and
antiquark by a colour flux string. It is possible to  arrange this
scheme so that one can define  modified {\it propagators} rather than
modified {\it operators}. This will have considerable  computational
advantages: once the appropriate fuzzed  propagator is defined, the
construction of hadronic correlations  is as fast as for local
propagators.

\begin{figure}[htb]

\vspace{16cm}

\includegraphics{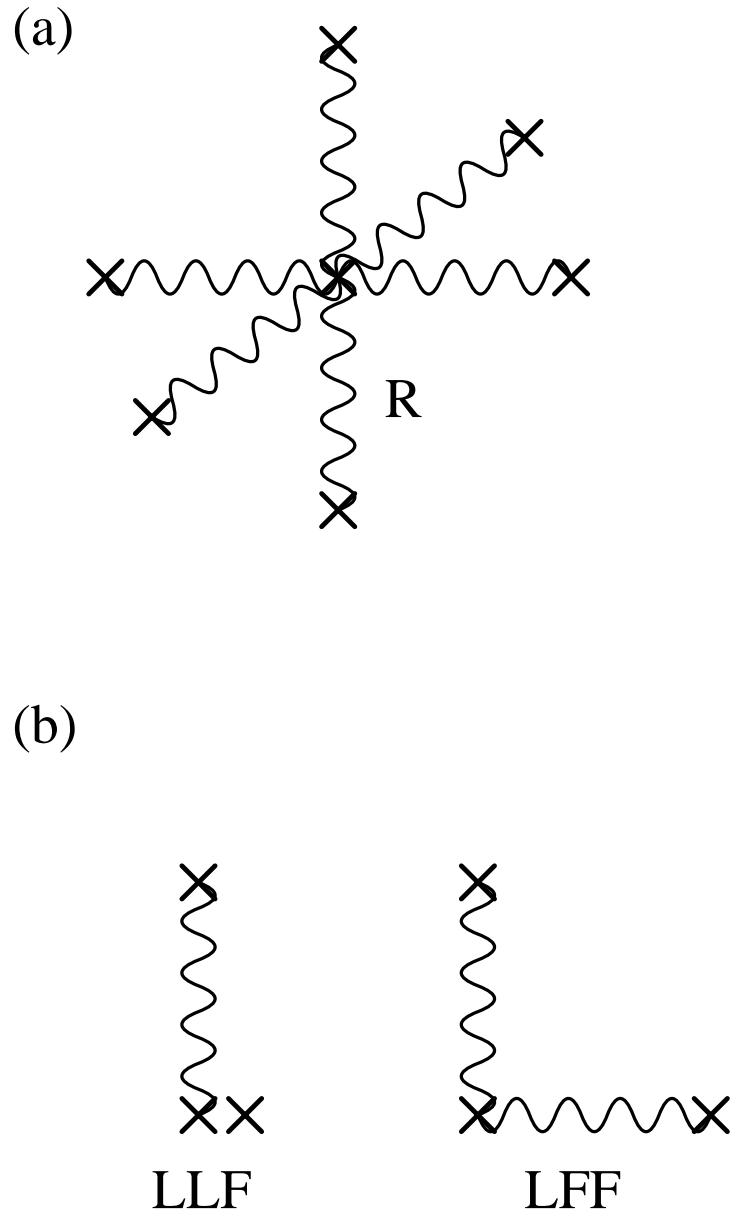}
\caption{
 (a) The spatially extended source used for the fuzzed  propagator. It
consists of a sum of  fuzzed links of length $R$ in the six spatial
orientations used.  
  (b) The  operators for baryons consisting of one fuzzed link
(LLF) and two fuzzed links (LFF) joining the quarks. A sum over 
the six spatial orientations of each fuzzed link is used.}
 \label{fig:fuzz}
\end{figure}

We can define a fuzzed quark propagator to a site $({\bf x},t)$ as the
average of the propagators to the sites given by the six spatial
displacements of distance $\pm R$ from $({\bf x}, t)$ along the lattice
axes together with the appropriate fuzzed links.  The sum over all six
orientations (three forward and three backward spatial directions)
results in an isotropic spatial dependence (in the spatial cubic group)
and so does not affect  $J^{PC}$ assignments. This is illustrated in
Fig.~\ref{fig:fuzz}(a), where the fuzzed links of length $R$ 
originating from the site $({\bf x},t)$ along the six orientations are
shown schematically. Then if such a fuzzed propagator is contracted
using an  appropriate $\gamma$ matrix  with a purely local  propagator
to the site $({\bf x}, t)$, a mesonic operator  is constructed with the
required properties.  We refer to this combination  as the LF operator.
Note that contracting two such fuzzed propagators (i.e.\ FF) with the
same value of $R$ is not likely to  be useful since  the fuzzed links
will partly cancel to the identity,  giving a component which is like a
purely local hadronic operator (i.e.\ LL). For baryons, three
propagators need to be combined. We find that  both LLF and LFF
combinations are useful. They are illustrated in 
Fig.~\ref{fig:fuzz}(b).

Since we have access to light quark propagators from a single  source
({\bf 0},0), we use the local operator at the source. At the sink at
time $t$, we use the spatially-extended  fuzzed operator of length $R$
described above. For notation we specify the source operator and then
the sink  operator used.  For the mesons we thus form local-fuzzed
(LL,LF) correlations by  replacing one local propagator in the usual
(LL,LL) formalism by the propagator fuzzed at the sink. 

For baryons we consider two different non-local operator correlations.
As shown in Fig.~\ref{fig:fuzz}(b),  these involve a di-quark separated
from a quark (single fuzzed - (LLL,LLF)) and an arrangement of three
quarks all separated (double fuzzed --- (LLL,LFF)). A triple fuzzed
configuration could also be studied but, by the same reasoning as for  
the mesons, it has a component which is unfuzzed and so gives poor
results. We also explored nucleon operators made of two fuzzed 
links at fixed angles ($0^0, 90^0$ and $180^0$). These operators 
are very computationally intensive to evaluate and gave no 
significant improvement in ground state extraction.

As remarked upon in Sec.\ 1, the method for constructing effective
operators described here has a direct physical interpretation only for
heavy-quark mesons. However, it will be  shown below that it can also be
successfully used for hadrons consisting of light quarks.

\section{Lattices and Fit Results}

We use light-quark propagators in the $24^3 \times 48$ configurations 
at $\beta=6.2$ obtained by UKQCD~\cite{ukqcd} using the clover improved
fermion action. Here we use values for the hopping parameter $K$ equal
to 0.14144 and 0.14262, which are the smallest/largest values
respectively considered in~\cite{ukqcd}. For the detailed investigation
we use  $K=0.14144$ (which corresponds to the largest quark mass), while
$K=0.14262$ is primarily used to investigate the quark mass dependence
of the results for the wave function. For both $K$ values the  most
comprehensive wave function results come from  an analysis of 12
configurations. We are able to supplement these detailed  wave function
measurements with correlation data at  fixed separation from  a larger
sample of 60 configurations~\cite{ukqcd,ukqcdnew}. This lattice study
corresponds to lattice spacing $a^{-1}=2.73(5) $GeV (determined from the
string tension), while the ratio $m_{\pi} / m_{\rho} \approx$ 0.77 and
0.52 for the two values of the hopping parameter respectively. Also, at
the smaller hopping parameter, $m_{\rho} \approx 1$ GeV, which is  close
to the value of the  physical state ($\phi$ meson)  built from  strange
quarks.

Using the effective operators described in the previous section, we can
now define the gauge-invariant correlation function (LL,LF) 
for mesons by 
 \begin{equation}
C(t,R)= \, <0| \overline{q}({\bf 0},0)
\Gamma q( {\bf 0},0) \frac {1}{6}
\sum_{ \bf x} \sum_{i=1}^6 \overline{q}({\bf x}+R_i,t) \Gamma M
  ({\bf x},t;R_i)  q({\bf x},t)|0>
\end{equation}
 where $M ({\bf x},t;R_i)$ denotes the product of fuzzed links of length
$R$ originating from site $\bf x$ in direction $i$  for a given time
slice $t$. The sum over spatial sites is needed to have momentum
zero, while the sum over the six directions is for the correct $J^{PC}$.
For baryons the expression is         similar to the one above, with the
$q \overline{q}$ component  suitably replaced by $ qqq$. The choice of
$\Gamma$ depends on the hadronic observable, e.g.\ $\Gamma = \gamma_5$
for the $\pi$ and   $ \Gamma = \gamma_i$ for the $\rho$. The explicit
definitions of the hadronic interpolating fields used here can be found
in~\cite{ukqcdpot}. The vector meson ($\rho)$ is obtained by averaging
over the three polarisation states, while for the nucleon we average the
$1,1$ and $2,2$  spinor indices of the correlator in the forwards time
direction ($ t < L_t/2$, where $L_t= 48$) and the $3,3$ and $4,4$
indices in the backwards time direction ($ L_t/2 < t \le L_t)$. For the
$\Delta$ one has additionally to project out the spin 3/2     
component.

Let us focus on accurate determinations of the ground state mass.  The
data  are usually presented by computing an effective mass. The  ground
state contribution dominates $m_{\rm eff}(t)=-\log(C(t,R)/C(t-1,R))$ at
large $t$ since  $m_0 < m_i$ for all $i \ne 0$.  In practice, extracting
the  ground state mass $m_0$ from data with statistical errors which
increase  with $t$ is subtle.  A good procedure is:
 
(i) a multi-exponential fit to the  widest acceptable $t$-range with

(ii) several hadronic operators (e.g.\ $R$-values) to stabilise the fit.

 One way to understand this guide is that the ground state is only 
determined accurately when an estimate of the first excited state  is
available. This is necessary since the energy difference controls the
rate  of approach of $C(t,R)$ to the expression given by the ground
state   component alone. However, fitting two (or more) exponentials to
just  one function  $C(t)$ is not very stable: it is better to have
several such functions (provided that they do indeed have different 
relative amounts of ground state and excited state).

\begin{figure}[htb]
\vspace{16cm}
\includegraphics{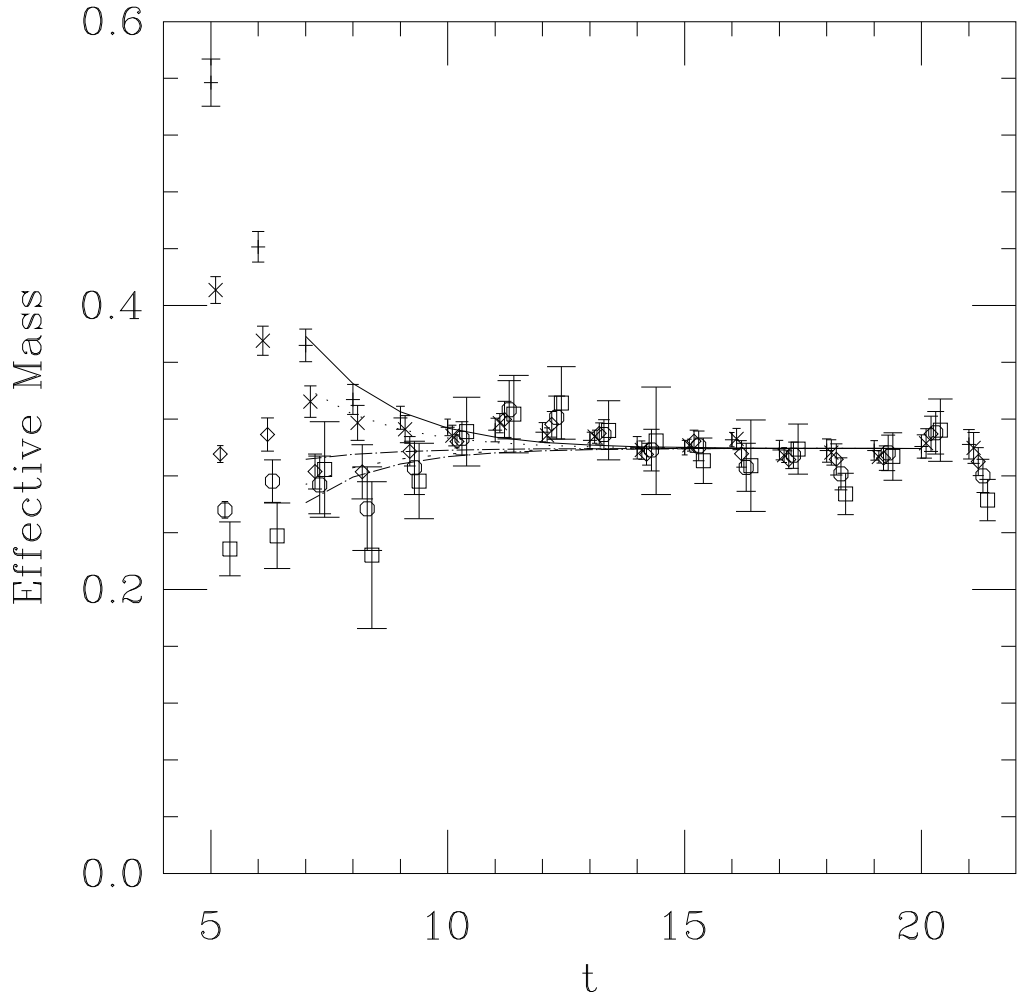}
\caption{
 The effective mass for the $\pi$ meson (in lattice units) using the
operators (LL,LL) corresponding to $R=0$ (+), and (LL,LF) with
R=4 ($\times$), 8 ($\diamond$), 10 ($\ast$) and 12 ($\Box$) at
$K=0.14144$.}
 \label{fig:pimass}
\end{figure}

\begin{figure}[htb]
\vspace{16cm}
\includegraphics{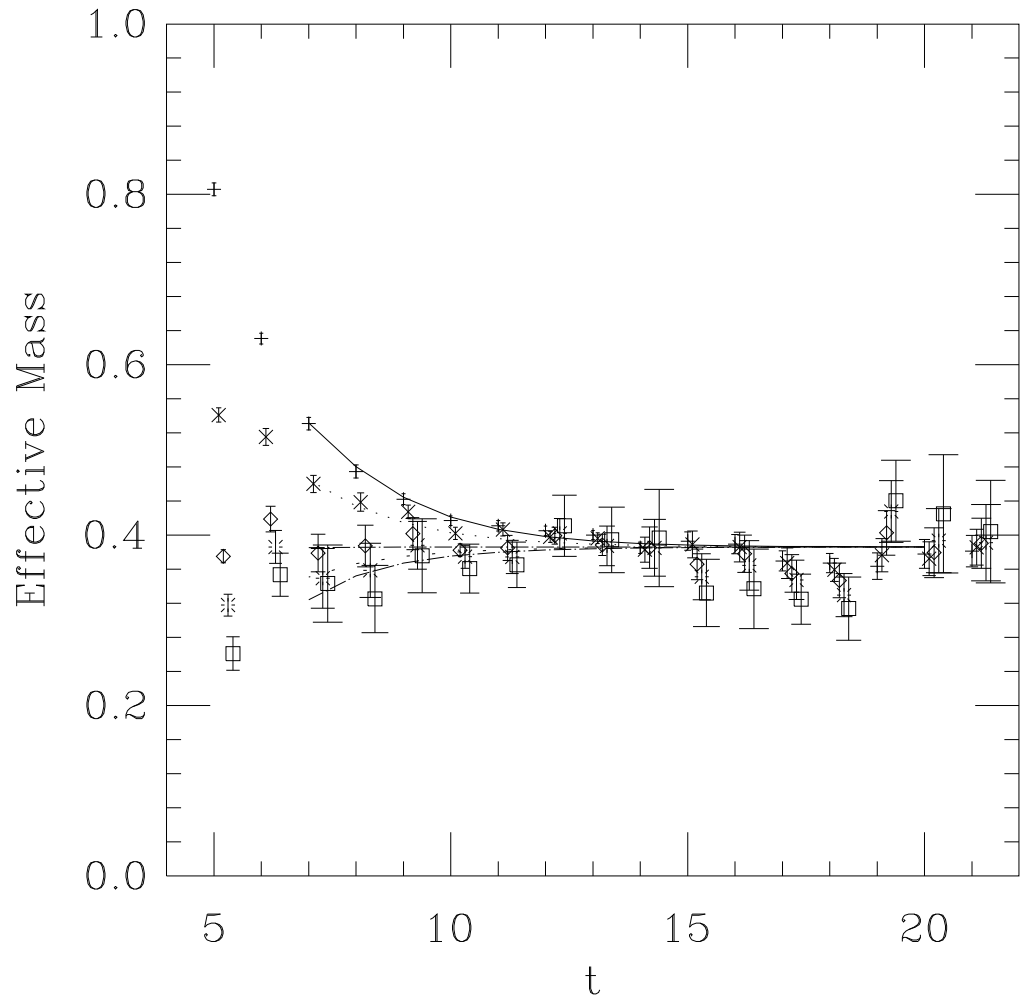}
\caption{
 The effective mass as in Fig.~\protect\ref{fig:pimass}, but for the
$\rho$ meson.}
 \label{fig:rhomass}
\end{figure}

\begin{figure}[htb]
\vspace{16cm}
\includegraphics{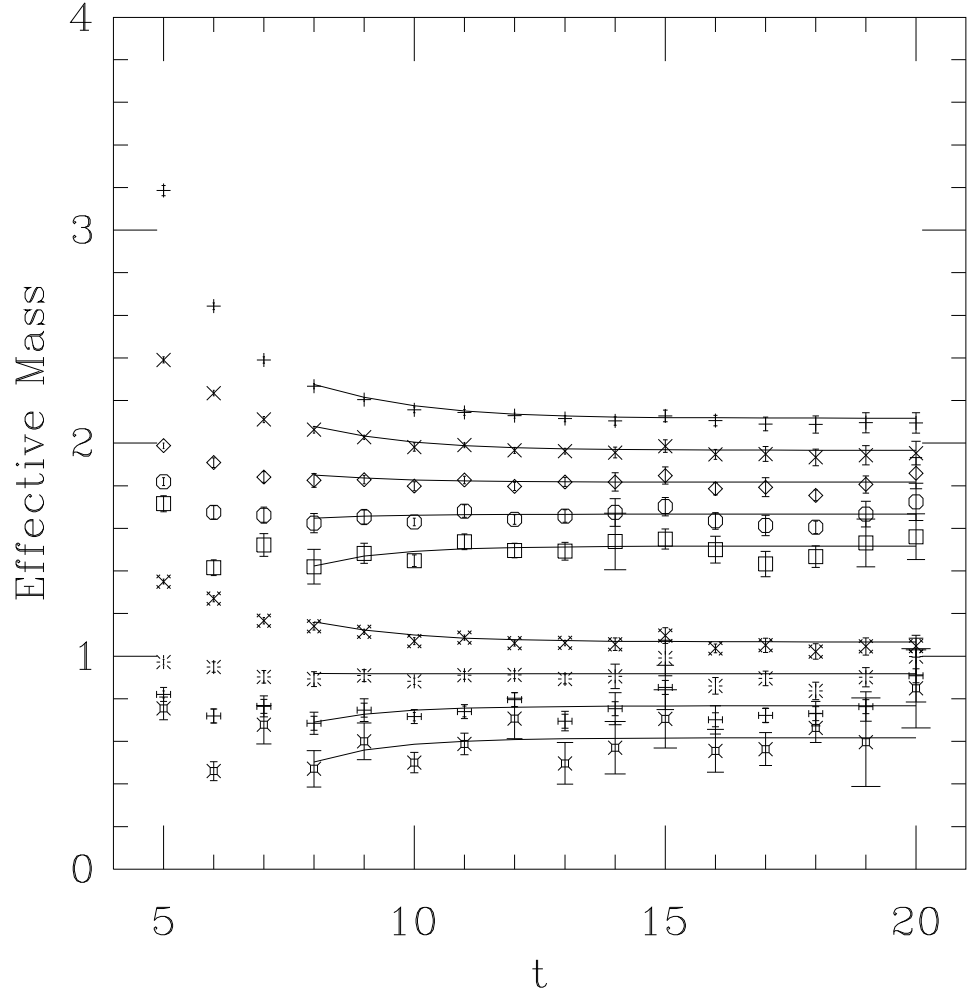}
\caption{
 The effective mass (in lattice units) for the nucleon at $K=0.14144$.
Here we show, from top to bottom, the (LLL,LLL) results corresponding to
$R=0$, (LLL,LLF) with $R$ = 4, 8, 10 and 12 and (LLL,LFF) likewise. In 
each case the plateau value of the curves at large $t$ is the same: the
results  have been displaced by a constant in the vertical direction for
legibility.}
 \label{fig:nuclmass}
\end{figure}

Thus we consider  fits to a two-exponential function
 \begin{equation}
C(t,R) = 
 c_0(R) e^{-m_0 t} + c_1(R) e^{-m_1 t} 
\label{fitf}
\end{equation}
 where periodicity in the time direction is understood in the case of
mesons where appropriate. We use a simultaneous fit to data at all $R$
values and $t$ values  by making use of correlated $\chi^2$ fits. When
dealing with a limited number of data samples,  conventional correlated
fits are known to give biased results~\cite{cor}. This is because  the
method for determining correlated $\chi^2$ is unreliable if the number
of data samples is insufficient (compared to the number of fit
parameters). Various methods have been suggested to deal with this. Here
we use the method proposed in~\cite{cmam}, which models the correlation
matrix by treating its inverse as 5-diagonal. The 2-exponential fits are
illustrated in Figs.~\ref{fig:pimass}--\ref{fig:nuclmass}. 

In the following, we use $R = 4, 8, 10$ and 12, which provide a broad
enough  spatial range for the hadrons considered here, as will be shown
below. The purely local operators (i.e.\ those containing only  LL and
LLL propagators) then correspond to $R = 0$. Statistical errors are
determined by means of a bootstrap analysis. Since we have only a small
data set at our disposal, the bootstrap samples tend to have even
stronger  correlations among the data. This can give anomalous error
estimates. We have, therefore, checked   the statistical errors using a
jackknife method. Even though this procedure is limited by the small
data size (the largest number of single-elimination jackknife blocks
being equal to the number of configurations, namely 12), we found the
errors to be of similar magnitude to those obtained from  the bootstrap
analysis.

The effective mass  $m_{\rm eff}(t)$  at $K=0.14144$ for the
pseudoscalar and vector mesons is shown in Figs.~\ref{fig:pimass} and
\ref{fig:rhomass}, and for the nucleon in Fig.~\ref{fig:nuclmass}.  For
the baryon,  where there are nine observables, the results for the
effective mass are shown spread out on the vertical axis to aid
legibility. The lines included in the figures are the results of the
two-exponential fit Eq.(\ref{fitf}). The results for the $\Delta$  are
very similar to those illustrated for the nucleon.

For both the mesons and baryons the results clearly show that, by using
fuzzed non-local operators, the plateau in the  effective mass as a
function of the time extent $t$   sets in at smaller values of $t$ than
is the case for the purely local observables.  Furthermore, the plateau
at the largest $R$ values tends to be approached  from below as $t$
increases. The fuzzed data have larger errors at lower $t$ (compared to
the unfuzzed ones), but for   larger $t$ the errors are comparable. The
local (unfuzzed) measurement thus appears to have   an accurate but
irrelevant component.

Although we have emphasized the advantages of making a simultaneous fit 
to several hadronic correlations, it is worth while to make a  direct
comparison of fits to either local operator data alone  or to fuzzed
operator data alone.  This will enable a comparison  to be made of the
effectiveness of each method for extracting  the ground state mass. Thus
we make single exponential fits in each case --- looking for a plateau
in $m_{\rm eff}$.  In Table~1 we list the results for the $\rho$ meson
of such fits to the purely local (LL,LL) and to the fuzzed (LL,LF)
correlation function with $R=8$.  At a given value of $t_{\rm min}$ the
errors for the ground state mass are smaller for the local operator.
However, if  one chooses $t_{\rm min}$ as small as possible consistent
with the $\chi^2$ per degree of freedom being acceptable (i.e.\ $ \leq
1$), the   errors on $m_0$ obtained from the fuzzed operators are
smaller.  Furthermore, the local operator fit shows that the value of
$m_0$  increases rapidly as $t_{\rm min}$ is reduced below 11. This
implies  that the results are very sensitive to an accurate estimate of 
$\chi^2$. The fuzzed operator fit remains stable to such a  reduction of
$t_{\rm min}$. Since, with highly correlated data,  the estimate of
goodness of fit can be unreliable, the sensitivity  of the local
operator fit to $t_{\rm min}$ is an additional systematic  error in that
case.

\section{Wave function}

The second aim of this paper is to study the hadronic (Bethe-Salpeter)
wave function determined from the non-local gauge-invariant operators
discussed in the previous section. Since we are limited by small
statistics for the fuzzed operator measurements, we choose  to
supplement our analysis by  using all available data to  fix the energy 
difference between the ground state and first excited state masses. In
particular, we make use of the full UKQCD data set of 60 
configurations~\cite{ukqcd} for the local operators (LL,LL) as well as
recent smeared results~\cite{ukqcdnew}. These consist of hadronic
correlations with quark propagators smeared   at the source and sink
(SS,SS) and at the source only (SS,LL)  by applying the Jacobi smearing
method  at $K=0.14144$. For the $\pi$ we have data with $\Gamma =
\gamma_5$ as well as $\gamma_4 \gamma_5$ at each end, so that we
effectively have twelve observables.

A big advantage of having smeared and local operators for the same
hadronic interpolating field  is that it allows a factorising
fit~\cite{fitxx}. These in turn provide tight  constraints on the ground
and excited state masses. We make 2-exponential fits to the  widest
$t$-range that gives an acceptable $\chi^2$. Correlations among  data at
different $t$-values are taken into account in the fit using several 
stable models~\cite{cmam}. These mass differences between the first
excited state  and the ground state  are then used in our wave function
extraction fit. For each hadronic channel, we then  fit simultaneously
all the local and non-local hadron measurements calculated from the
subset  of 12 configurations to the  fit function Eq.~(\ref{fitf}). The
coefficients $c_0(R)$ and $c_1(R)$  are then the required wave function.
The values obtained from  the fit are normalised so that
$c_0(0)+c_1(0)=1$ at $t=t_{\rm min}$. 

\begin{figure}[htb]
\vspace{7cm}
\includegraphics{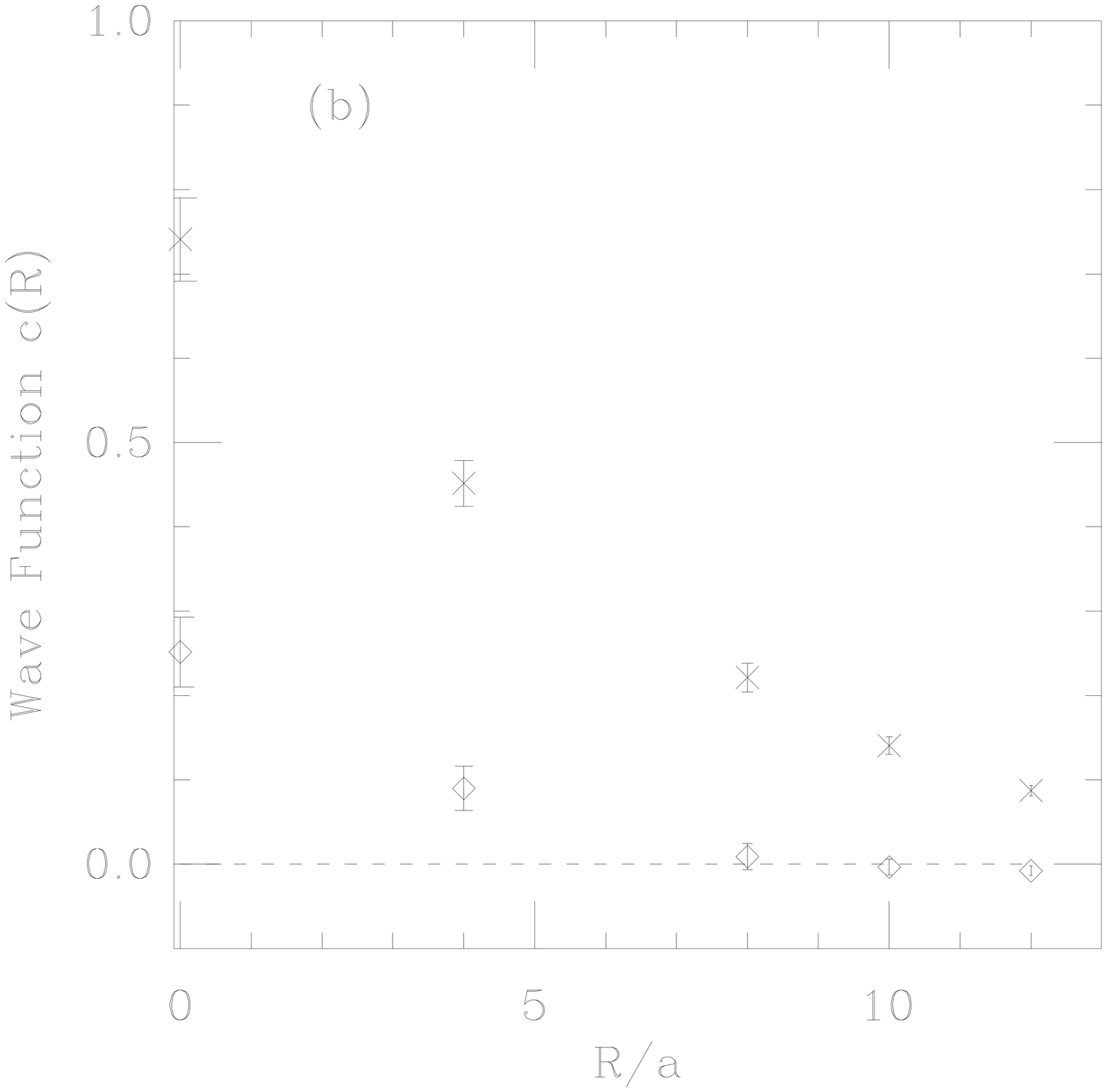}
\caption{
 The wave function for the ground state ($\times$) and first excited
state ($\diamond$) for (a) the $\pi$ meson and (b) the $\rho$  meson at
$K=0.14144$.}   
 \label{fig:wfpirho}
\end{figure}

\begin{figure}[htb]
\vspace{7cm}
\includegraphics{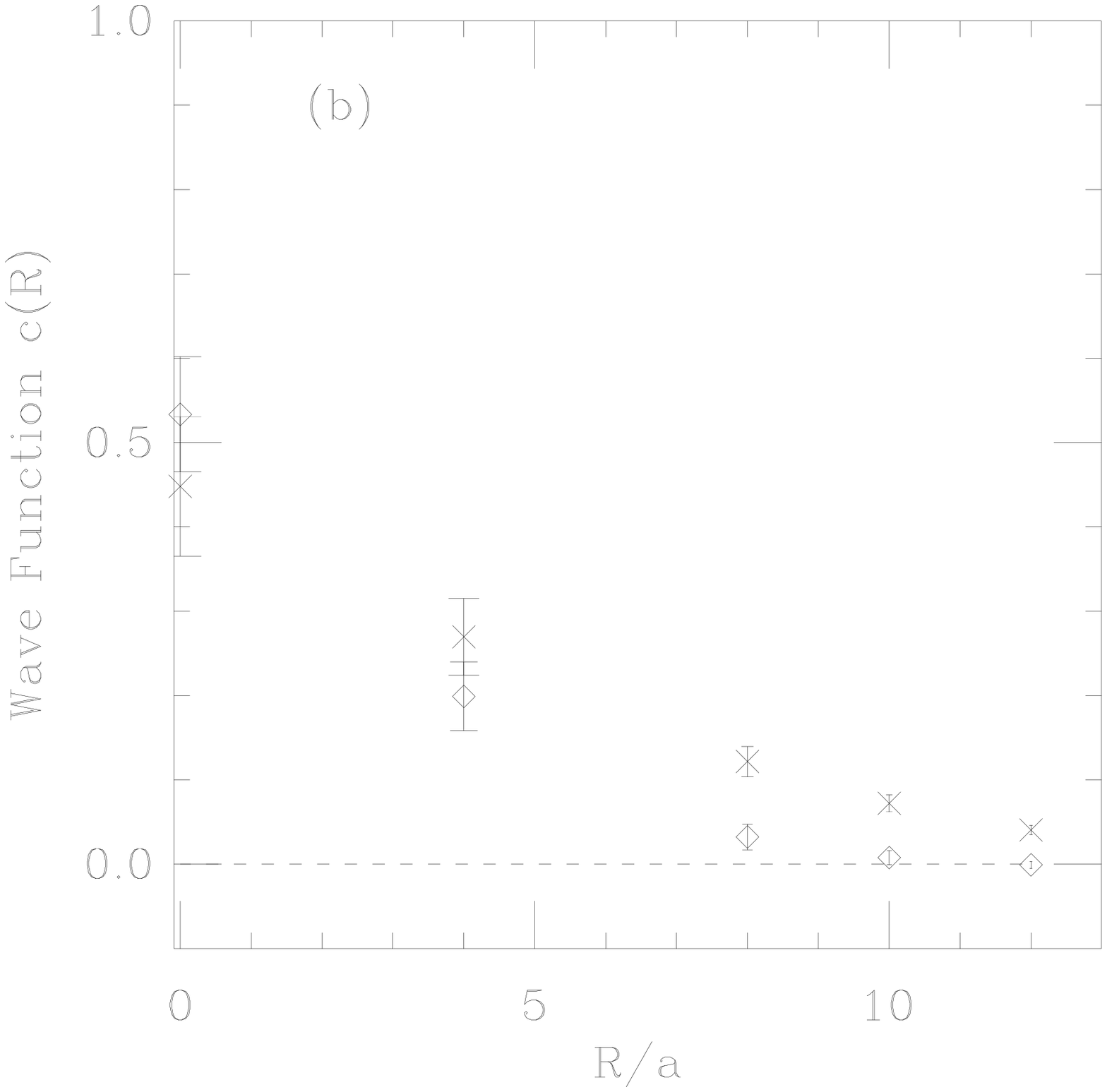}
\caption{ 
 The wave function from LLF operators for the ground state ($\times$)
and first excited state ($\diamond$) for the nucleon (a) and $\Delta$
(b) at $K=0.14144$. Also shown for the nucleon are the results at
$K=0.14262$ for the ground state ($\Box$) and first excited state
($\ast$).}   
 \label{fig:wfbaryons}
\end{figure}

The wave function for the ground and first excited states obtained from
the local operators and those involving only the fuzzed link of length
$R$, normalised as outlined above, are shown in Fig.~\ref{fig:wfpirho}
for the mesons and in Fig.~\ref{fig:wfbaryons} for the baryons at
$K=0.14144$. The behaviour of the ground state wave function for mesons
in quenched QCD has recently been discussed in detail (cf.
e.g.~\cite{mesonwf,mesonwf1, bs-gup,tn}). Our results are  in agreement
with those obtained in the literature using similar   gauge-invariant
definitions of the wave function~\cite{bs-gup,tn}.

As far as we are aware, the excited state wave function has     not been
studied either for mesons or baryons. The interesting feature that can
clearly be seen for all the    hadronic observables considered here is
the presence of a node in the excited state wave function as a function
of the length of fuzzed links  connecting the quark and antiquark at the
sink. At this $R$-value,  the ratio of ground state wave function to
excited state wave function becomes zero. This particular spatial extent
($R \approx 8$), which is more or less the same for all the observables,
thus seems to be an optimal choice for determining e.g.\ the ground
state mass, since the  contamination of the ground state by higher
excited states has been minimised. At larger $R$-values, the effective
mass  plateau is seen to be reached from below as $t$ increases. This is
explained by the  change in sign of the excited state wave function.

To investigate what happens if the quark mass decreases, we also show in
Fig.~\ref{fig:wfbaryons}(a) the results for  the nucleon wave function
at $K=0.14262$. Although we only consider $R=0$ and $8$, the results are
in agreement with those at the smaller $K$ value at the same distance.
This  is consistent with earlier conclusions regarding the independence
of the ground state wave function on the value of the  quark
mass~\cite{mesonwf1,bs-gup}.   Although the interpretation of the wave
function as  an indication of the physical size of the (heavy) hadron 
loses its meaning if the quark mass is decreased and our physical
motivation is no longer strictly applicable, the operators described
above are still effective in producing a clean ground state. This gives
us confidence that these operators can be used regardless of the value
of the hopping parameter (quark mass).

\begin{figure}[t]

\vspace{7cm}

\includegraphics{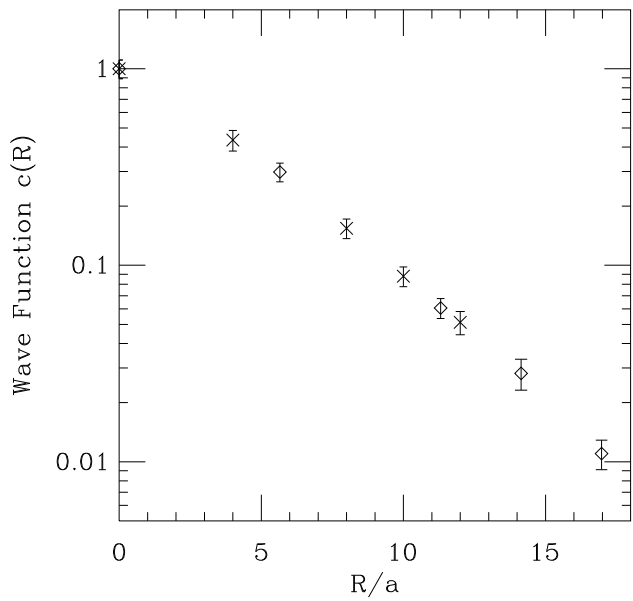}
\caption{
 The nucleon ground state wave function, using the (LLL,LLF)($\times$)
and (LLL,LFF)($\diamond$) operators at  $K=0.14144$   replacing $R$ by 
$ \protect\sqrt{2} R$ for the (LLL,LFF) results.}  
 \label{fig:wfnuclgs}
\end{figure}

For the nucleon and $\Delta$, where we have two different non-local
operators at our disposal, the ground state wave function for the LLF
operator is broader than that for the LFF one. This can be understood 
since, for a given $R$ value, the quarks are on average farther apart in
the LFF case than  the LLF case, see Fig.~\ref{fig:fuzz}(b). Thus, if in
the LFF case we replace  $ R \rightarrow {\sqrt 2} R$, we see in
Fig.~\ref{fig:wfnuclgs} that the results  for the LLF and LFF operators
lie on a single curve. This behaviour is in agreement with what one
would naively expect: the colour charges at the end points of the LFF
baryon operator at the sink are effectively ${\sqrt 2} R$  apart and so
should be compared with  the LLF wave function at ${\sqrt 2} R$. 
Further investigation of the angular dependence of the baryon wave
function is possible using operators with two fuzzed links at
fixed angles to each other. Because these operators are not obtainable
from our fuzzed  propagator construction, they involve two orders of
magnitude more  computation than the LFF operator which is a sum over a
specific  combination of them. 

\begin{figure}[t]

\vspace{6cm}

\includegraphics{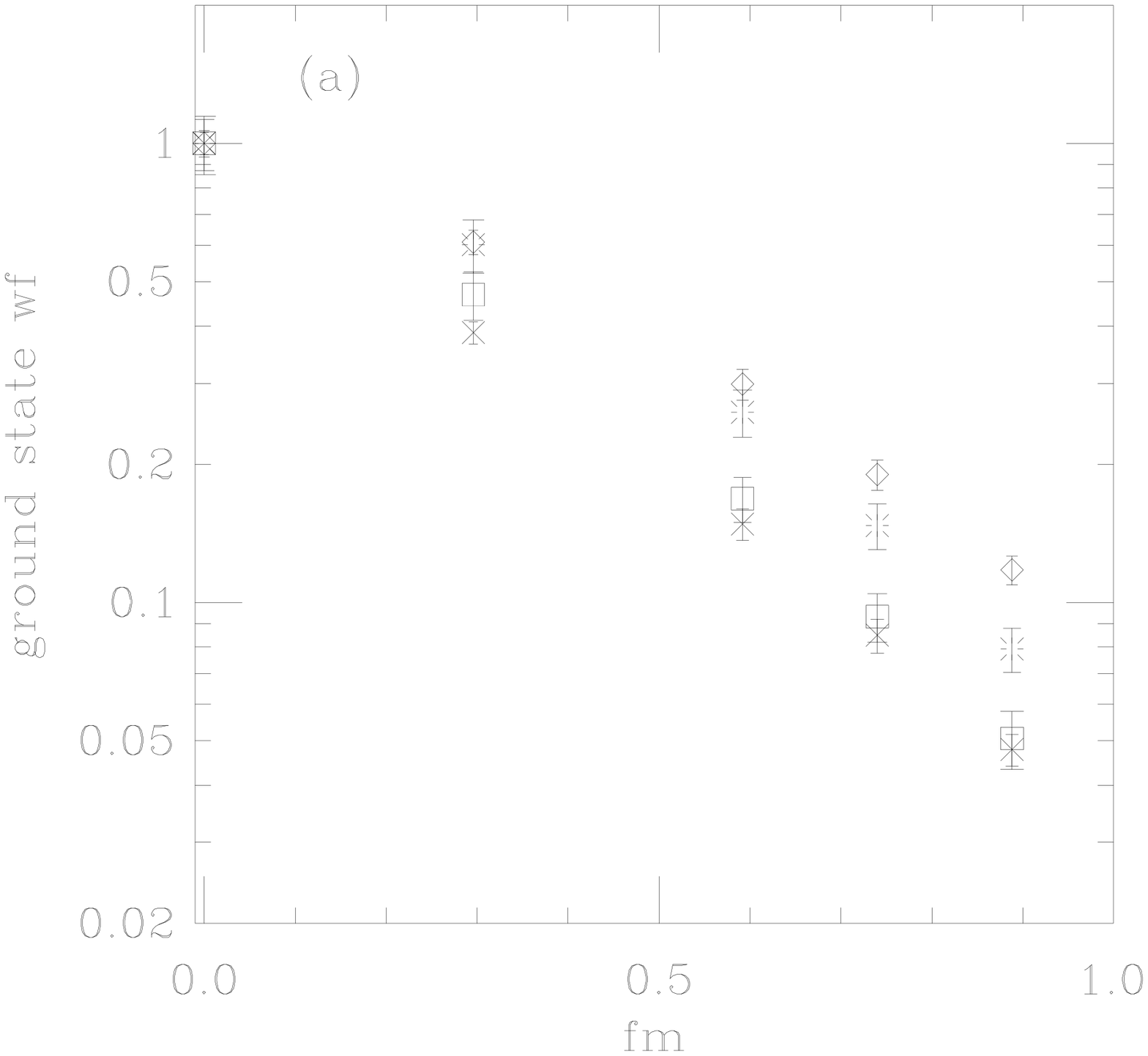}
%\end{figure}

%\begin{figure}[htb]

\vspace{9cm}

\includegraphics{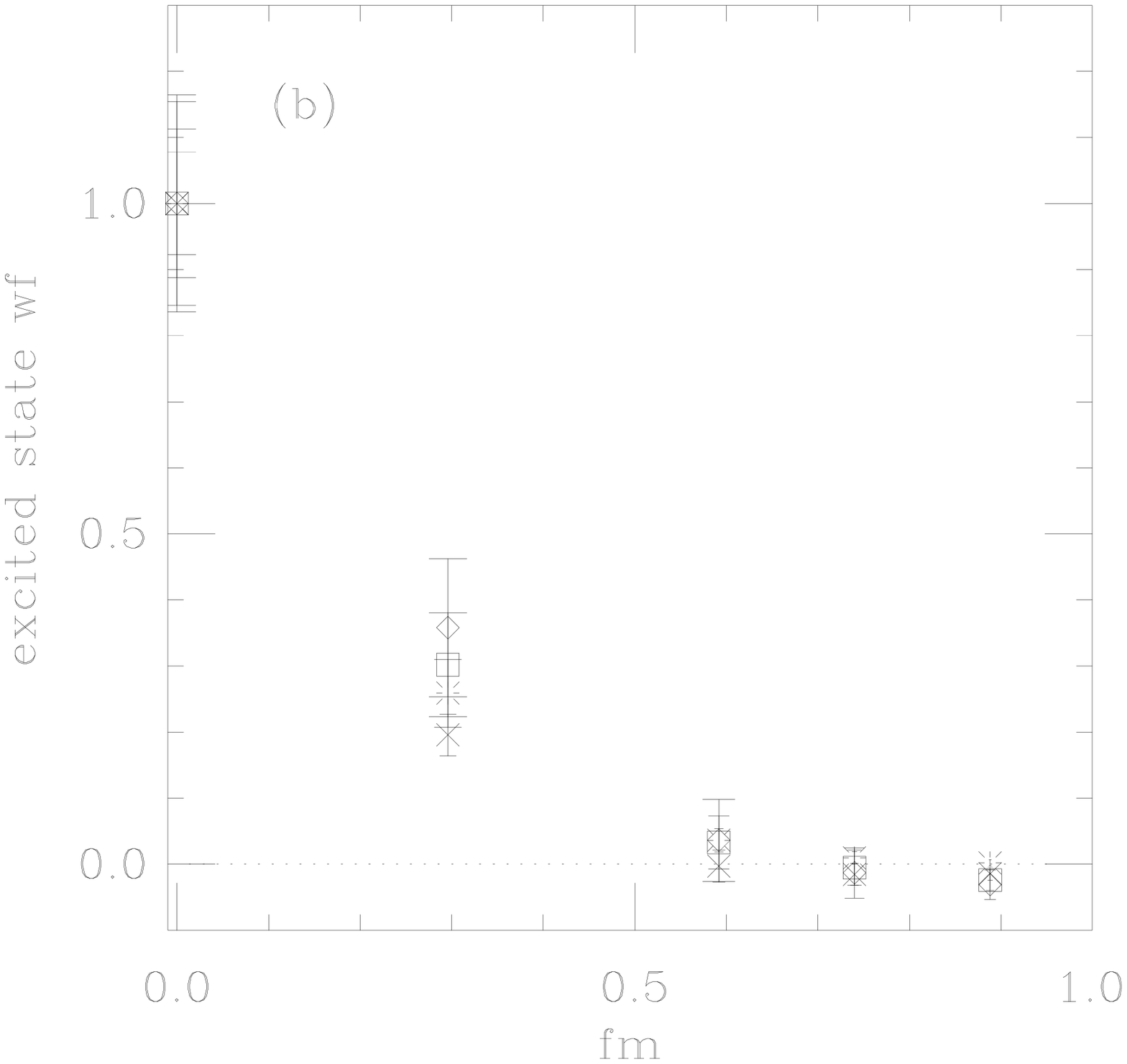}
\caption{
 (a) The ground state  and (b) excited state  wave function for the $\pi$
($\times$), $\rho$ ($\diamond$), nucleon ($\Box$) and $\Delta$ ($\ast$)
at $K=0.14144$ as a function of the physical size in fm. The values are
normalised to one at distance $R$=0.}  
 \label{fig:wfmesbar}
\end{figure}

The relative sizes of the hadronic wave function, normalised to one at
zero distance, are shown in Figs.~\ref{fig:wfmesbar}(a) and
\ref{fig:wfmesbar}(b) as functions of the  physical distance for the
ground states and excited states respectively. For the baryons we use
the LLF operators. To obtain the distance scale in physical units we use
the value of the inverse lattice spacing determined by UKQCD from the
static  potential [$ a^{-1} \approx 2.7 $ GeV]. The behaviour for the
ground state wave function  is as expected, with the $\rho$ meson and
$\Delta$ the largest, while the $\pi$ meson  and nucleon show similar
sizes ~\cite{mesonwf,mesonwf1}. The results for the excited state wave
function for the different hadrons seem to be in agreement, even though
the error bars are too    large to make a definite claim. In each case
the results are consistent with a node at $R \approx  3$ GeV$^{-1}$.

In our investigation above we have used a fixed fuzzing level  and
corresponding value of $c$ (see Eq. (\ref{eqf})). We have, however,
checked that the effective mass remains  essentially unchanged if a
higher fuzzing level is used. It is known from  earlier studies of the
wave function that fuzzing of the gauge links is vital to obtain a good
overlap with the hadronic state one is studying. It was found 
in~\cite{bs-gup}, where a slightly different fuzzing algorithm was used,
that the result for the wave function as a  function of the fuzzing
level converges rapidly as the level approaches 6 (in their method).
This would roughly correspond to the level used here. Hence we do not
expect the results presented above to change in any significant way if
the fuzzing level were to be increased.

The computational overhead needed to construct the fuzzed propagator  at
the sink from a local propagator is quite small. With fuzzing level  $f$
(we used $f=5$), then approximately $30 f +96$ matrix multiplications 
are needed per site ($3 \times 3 $ complex matrix multiplications) to 
create the fuzzed propagators.  For the Jacobi smearing method with  $j$
iterations (where $j=50$ is typical),  approximately $96j$ matrix
multiplications per site are needed. Thus the computational overhead is
considerably  smaller in the fuzzed prescription. In essence this comes
from  treating the sink/source as a sum of just 6 points for the fuzzing
method. At the source, there is essentially an equal computational
overhead in  either case since smearing/fuzzing is only needed  around
one site.

One drawback of the fuzzed propagator approach is that FF hadronic
operators  are not useful so one needs to combine local propagators to
construct  LL and LF combinations. For the Jacobi smearing, on the other
hand, SS  operators are useful. In general, however, it helps to have at
least two operator constructs in use so that multi-exponential fits  can
be stabilised. This implies that Jacobi smearing should also  make use
of SL operators if possible. Thus each method is improved if  both local
and smeared/fuzzed  propagators are  available.

\section{Conclusions}

We have shown how to define a fuzzed quark propagator in an efficient 
way. Combined with the usual local propagator, this then enables 
hadronic propagation to be studied with a wave function motivated  by
the heavy quark picture. This hadronic operator of length $R$  can be
tuned to have no contribution from the first excited state  by varying
$R$. This gives a plateau in the effective mass which extends to  lower
$t$-values. Equivalently, the measured Greens function will  be
dominated by the ground state at smaller $t$ --- and so will  make 
calculations of ground state matrix elements more reliable.  Our
prescription for creating extended hadronic operators seems to  work
equally well for all hadrons with the same fuzzed quark  propagators as
ingredients. This makes it computationally efficient.

We have supported the view that 2-exponential fits to the $t$-dependence
 of hadronic correlations  are needed to extract reliable ground state 
masses. Such fits need several hadronic operators to make them  stable.
Choosing fuzzed operators with different separations $R$ is an 
attractive and efficient way to do this.

\section{Acknowlegdements} 
  
  This work was supported by SERC grants GR/G 32779, GR/J98202 and  
GR/H 01236  and EC grants CHRX-CT92-0051 and ERBSCI*CT91-0642.

\begin{table*}[hbt]

\setlength{\tabcolsep}{1.5pc}

\newlength{\digitwidth} \settowidth{\digitwidth}{\rm 0}
\catcode`?=\active \def?{\kern\digitwidth}

\caption{Comparison of single exponential fit results
 for the $\rho$ meson.}
\label{tab:effluents}
\begin{tabular*}{\textwidth}{@{}l@{\extracolsep{\fill}}rrrrr}
\hline
                 & \multicolumn{2}{l}{(LL,LL) Correlation} 
                 & \multicolumn{2}{l}{(LL,LF) Correlation} \\
\cline{2-3} \cline{4-5}
[$t_{\rm min},t_{\rm max}$] & $m_0$ & $\chi^2/{\rm dof}$ & $m_0$ & 
$\chi^2/{\rm dof}$ \\
\hline
$[5,20]$           & $~~  $ & $~~ $ & $ 0.3898(74)  $ & $ 0.8 $ \\
$[6,20]$          & $~~  $ & $~~ $ & $ 0.3863(74)  $ & $ 0.5 $ \\
$[7,20]$          & $~~  $ & $~~ $ & $ 0.3877(87)  $ & $ 0.6 $ \\
$[8,20]$          & $0.4212(122)  $ & $3.8 $ & $ 0.3891(97)  $ & $ 0.8 $ \\
$[9,20]$          & $0.4052(93)  $ & $1.2 $ & $ 0.3839(97)  $ & $ 0.9 $ \\
$[10,20]$          & $0.4004(92)  $ & $1.0 $ & $ 0.3841(109)  $ & $ 0.9 $ \\
$[11,20]$          & $0.3951(88)  $ & $0.8 $ & $ 0.3855(129)  $ & $ 0.9 $ \\
$[12,20]$          & $0.3907(88)  $ & $0.7 $ & $ 0.3789(142)  $ & $ 0.9 $ \\
\hline
\multicolumn{5}{@{}p{120mm}}{}
\end{tabular*}
\end{table*}

\newpage

\newpage

\newpage

\newpage

\newpage
\eject

\begin{figure}[htb]
\vspace{6cm}
\includegraphics{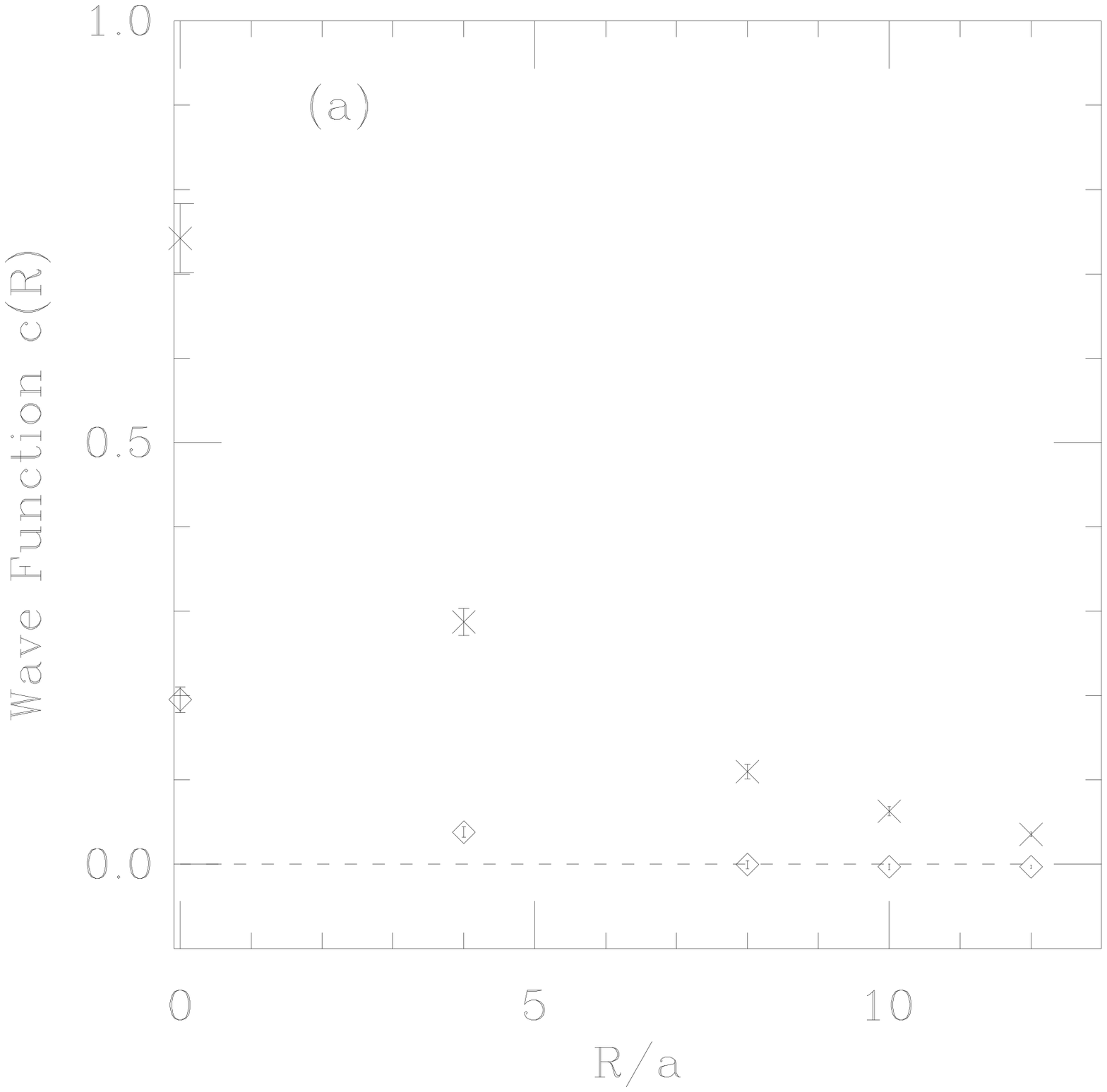}
\end{figure}

\newpage
\eject

\begin{figure}[t]
\vspace{6cm}
\includegraphics{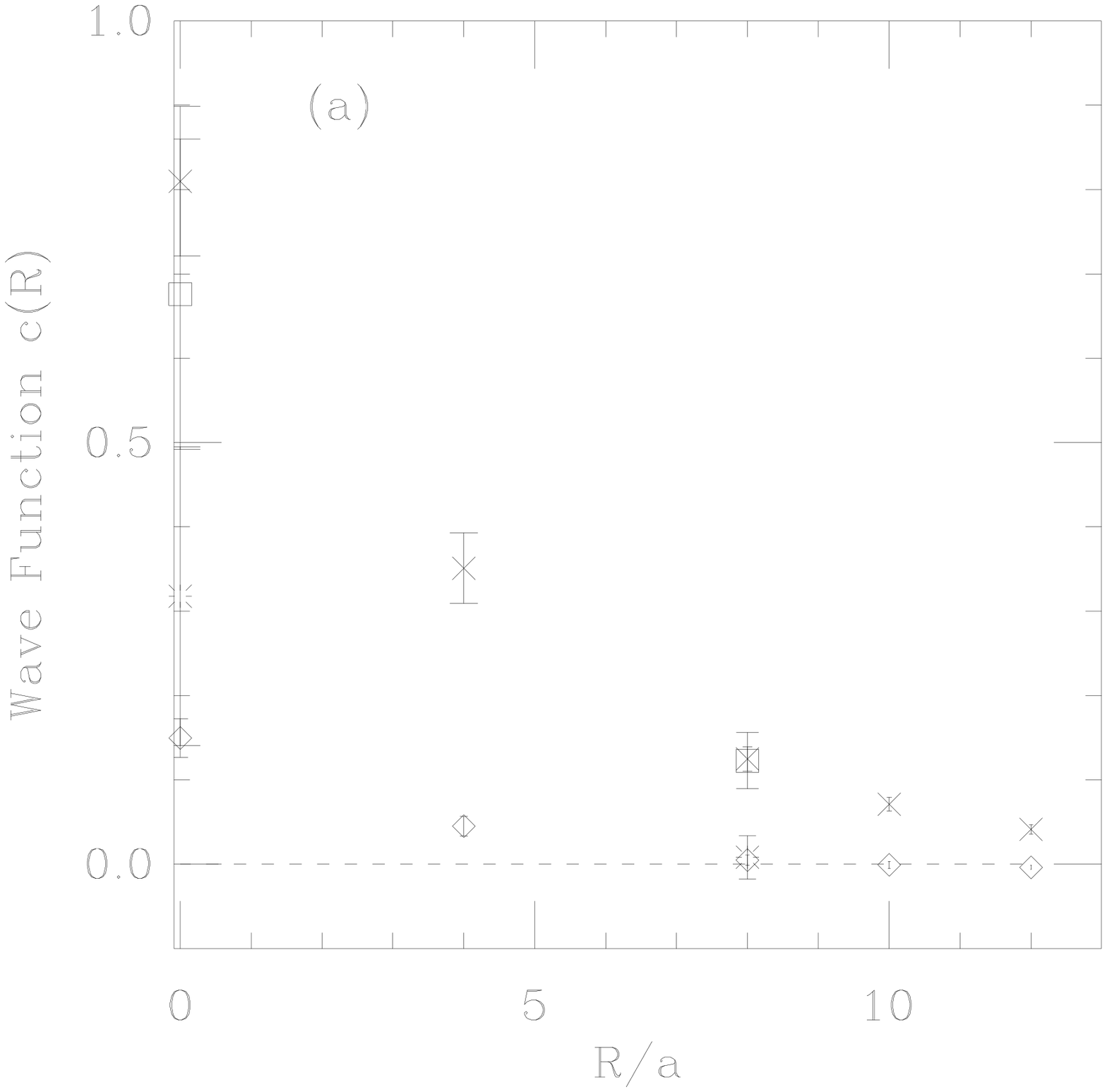}
\end{figure}

\newpage
\eject

%\begin{figure}[htb]
%\vspace{7cm}
%\special{psfile=fig8a.ps voffset=-30 hscale=45 vscale=45}

%\end{figure}

%\newpage
\vfill
\eject

\end{document}